\documentclass[sigconf]{acmart}
\AtBeginDocument{%
  \providecommand\BibTeX{{%
    \normalfont B\kern-0.5em{\scshape i\kern-0.25em b}\kern-0.8em\TeX}}}

\copyrightyear{2023} 
\acmYear{2023} 
\setcopyright{rightsretained} 
\acmConference[WWW '23 Companion]{Companion Proceedings of the ACM Web Conference 2023}{April 30-May 4, 2023}{Austin, TX, USA}
\acmBooktitle{Companion Proceedings of the ACM Web Conference 2023 (WWW '23 Companion), April 30-May 4, 2023, Austin, TX, USA}
\acmDOI{10.1145/3543873.3584654}
\acmISBN{978-1-4503-9419-2/23/04}

\usepackage{nicefrac}
\usepackage{numprint}
\usepackage{enumitem}
\usepackage{multirow}

\begin{document}

\title{Measuring e-Commerce Metric Changes in Online Experiments}
% \subtitle{Empirical Evidence on Measurement Uncertainty in Average Basket Value and Other e-Commerce KPIs}


\author{C. H. Bryan Liu}
\affiliation{%
  \institution{ASOS.com \& Imperial College London}
  \country{United Kingdom}}
\email{bryan.liu12@imperial.ac.uk}
\orcid{0000-0002-6516-2364}
\authornote{This work is supported by ASOS.com and the EPSRC Centre for Doctoral Training in Modern Statistics and Statistical Machine Learning (StatML.IO).}

\author{Emma J. McCoy}
\affiliation{%
  \institution{London School of Economics and Political Science}
  \country{United Kingdom}}
\orcid{0000-0001-9033-7472}

\renewcommand{\shortauthors}{Liu and McCoy}

\begin{abstract}
  Digital technology organizations routinely use online experiments (e.g. A/B tests) to guide their product and business decisions. In e-commerce, we often measure changes to transaction- or item-based business metrics such as Average Basket Value (ABV), Average Basket Size (ABS), and Average Selling Price (ASP); yet it remains a common pitfall to ignore the dependency between the value/size of transactions/items during experiment design and analysis. We present empirical evidence on such dependency, its impact on measurement uncertainty, and practical implications on A/B test outcomes if left unmitigated. By making the evidence available, we hope to drive awareness of the pitfall among experimenters in e-commerce and hence encourage the adoption of established mitigation approaches. We also share lessons learned when incorporating selected mitigation approaches into our experimentation analysis platform currently in production.\footnote{The experiment code and results on the two publicly available datasets are available on GitHub/Zenodo: \url{https://doi.org/10.5281/zenodo.7659092}.}
\end{abstract}





\maketitle

\section{Introduction}
\label{sec:abv_intro}

Online experiments have become popular among digital technology organizations in measuring the impact of their products/services and guiding business decisions~\cite{kohavi20trustworthy,liu20whatisthevalue}. The simplest example, an A/B test, randomly splits incoming users into a treatment and a control group. We calculate some decision metrics
based on responses from both groups and compare the metrics using a statistical test to draw causal
statements about the treatment.

% e-commerce: What's similar, what's different
Here we focus on A/B tests in e-commerce. Similar to many technology organizations, e-commerce organizations aim to provide a consistent user experience, and hence randomization in A/B tests is generally done on a per-user basis. On the other hand, e-commerce organizations are unique as they feature physical inventories and thus track a set of business metrics that are transaction- or item-based. These metrics include Average Basket Value (ABV),\footnote{Average (mean) amount spent in each transaction/basket by a user.} Average Basket Size (ABS),\footnote{Average (mean) number of items purchased in each transaction/basket by a user.} and Average Selling Price (ASP).\footnote{Average (mean) price per item sold.}

Experiments that measure changes to ABV, ABS, and ASP often feature dependent responses.
In these experiments, the responses (or \emph{analysis units}) are at a more granular transaction or item level than the \emph{randomization units}, which are at a user level. Given a user can make many transactions and purchase many items during an experiment, the value of these transactions and items are likely to be correlated based on the user's preference. This creates a local dependence structure between users and transactions/items, which violates the usual independent and identically distributed (i.i.d. hereafter) assumptions in common statistical tests that A/B tests employ (e.g., a Student's $t$-test). If left unmitigated, it risks experimenters having wrong estimates of sampling uncertainty and hence making wrong conclusions from a statistical test.

Having dependent responses in experiments due to differences in granularity between randomization and analysis units is not a recent revelation~\cite{kohavi20trustworthy}. In some sense, one may regard such a setup as a cluster-randomized controlled experiment, where an experimenter randomly assigns clusters of transactions or items belonging to the same user to the A/B test groups.
In terms of obtaining an unbiased estimate for the sampling uncertainty, many viable approaches already exist: experimenters in medicine, economics, and social sciences often employ a crossed random effects model~\cite{baayen08mixedeffects} or cluster robust standard errors~\cite{cameron11robust}. Meanwhile, those in digital technology prefer using bootstrap~\cite{bakshy13uncertainty} or the delta method~\cite{deng18applying} to estimate the sample variance and the standard error due to their ability to scale to large datasets and relative lack of model assumptions.
% \footnote{In such a setup, the size of a cluster can be zero and would only be determined after the experiment ends. This is due to experimenters not able to control how much a user buys on the website.} 

\begin{table*}
  \caption{Summary of the three online retail/e-commerce transactions datasets featured in this paper (post data cleaning).}
  \label{tab:ecomm_dataset_summary}
  \vspace*{-6pt}
  \begin{tabular}{cccccc}
    \toprule
    Dataset & \# Users/Customers & \# Transactions/Orders & \# Items/Units & \# Products/SKUs & Time span\\
    \midrule
    ASOS (Proprietary) &
    $\sim$\numprint{4180000} &
    $\sim$\numprint{9680000} &
    $\sim$\numprint{32900000} &
    $\sim$\numprint{330000} &
    62 days (2 months) \\
    UCI Online Retail II~\cite{dua2019UCI,chen12datamining} &
    \numprint{5852} &
    \numprint{36594} &
    \numprint{10690447} &
    \numprint{4621} & 
    739 days (2 years) \\
    Olist Brazilian e-Commerce~\cite{olist18braziliandataset} &
    \numprint{94983} &
    \numprint{98199} &
    \numprint{112101} &
    \numprint{32729} & 
    729 days (2 years)\\
    \bottomrule
  \end{tabular}
\end{table*}

% Gap particularly for e-commerce
Despite the number of work dedicated to dependent responses in digital experiments, there is little published evidence specifically in the context of A/B tests in e-commerce. Most of the work available is based on the context of A/B tests in digital advertising and content, where experimenters randomize by users and analyze by sessions or page views~\cite{bakshy13uncertainty,deng18applying}; or social networks, where experimenters both randomize and analyze by users whose responses may correlate via their social connections~\cite{eckles17design,gui15network}. We believe the lack of evidence contributes to insufficient awareness of the issue from experimenters in this area.

This paper's contribution is thus evidence from real-life e-commerce experimentation operations. In particular:

\begin{enumerate}[leftmargin=*]
    \item We show, using three e-commerce transaction datasets (two of which are publicly available), a positive correlation between the value/size of transactions and items from the same user. This leads to inflation in measurement uncertainty (Section 2);
    \item We quantify the extent of the said inflation, which can be anywhere between 1x to >100x the usual estimate and is dependent on the business metric, experiment duration, and organization (Section 2.1);
    \item We highlight the impact of the said inflation on test power and confidence interval coverage in null hypothesis significance tests (Section 2.2); and
    \item We share lessons learned while incorporating some of the established mitigation approaches into our experiment analysis platform, including a critique of some popular approaches and practical trade-offs during implementation (Section 3).
\end{enumerate}

\section{Beware of Hidden Inflation in Measurement Uncertainty}

% \subsection{Why is the standard error (SE) inflated?}
We first motivate why we may obtain inaccurate estimates of the sampling uncertainty when measuring changes to e-commerce metrics based on transactions or items in A/B tests. Let ${X_1, X_2, \dots, X_n}$ be our responses with $\mathbb{E}(X_i) = \mu$ and ${\textrm{Var}(X_i) = \sigma^2}$. The test statistic for a (one-sample) Student's~$t$-test, the most commonly used statistical test, is $(\bar{x} - \mu)/\sqrt{s^2/n}$,
where~$\bar{x}$ is the sample mean and~$s^2$ is the sample variance. The denominator of the test statistic is an unbiased estimate of the standard error of the mean (standard error, or SE hereafter):
\begin{align}
  \sqrt{\textrm{Var}(\bar{X})}
%   & = \sqrt{\textrm{Var}\left(\nicefrac{1}{n} \textstyle\sum_{i=1}^{n} X_i\right)} \nonumber\\
  & = \sqrt{\nicefrac{1}{n^2} \big(\textstyle\sum_{i=1}^{n} \textrm{Var}(X_i) + 2 \textstyle\sum_{i<j} \textrm{Cov}(X_i, X_j) \big)} \;.
\end{align}
If we assume $X_i$ are i.i.d., the prevailing assumption when experimenters randomize by users and analyze by user~\cite{deng17trustworthy}, then $\textrm{Cov}(X_i, X_j) = 0 \;\;\forall i \neq j$ and thus ${\textrm{Var}(\bar{X}) = \sum_i \textrm{Var}(X_i)/n^2 = \sigma^2/n}$.

The i.i.d. assumption is unlikely to hold when experimenters randomize by users and analyze by transactions or items. Many users tend to make a transaction that is similarly sized and valued to their previous transaction(s). They also tend to purchase items that are at a similar price point.
% Introduce the datasets
To demonstrate the claims above, we utilize three online retail/e-commerce transactions datasets. The UCI Online Retail II~\cite{chen12datamining,dua2019UCI} and Olist Brazilian e-Commerce~\cite{olist18braziliandataset} datasets are publicly available.\footnotemark[1] We also use a proprietary dataset from ASOS.com, a global fashion e-tail company, that records transactions within a specific two-month period in 2022 on a particular mobile platform.\footnote{The data described are not representative of ASOS.com's overall business operations and one should not draw any such conclusion from the dataset.} We summarize the three datasets in Table~\ref{tab:ecomm_dataset_summary}.

\begin{figure}
  \begin{center}
    \includegraphics[height=0.17\textwidth, trim = 0mm 2mm 2mm 2mm, clip]{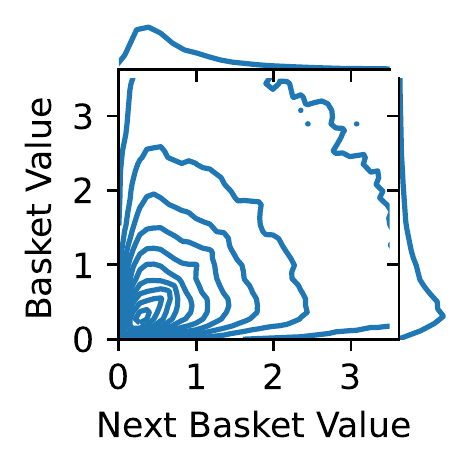}
    \includegraphics[height=0.17\textwidth, trim = 10mm 2mm 2mm 2mm, clip]{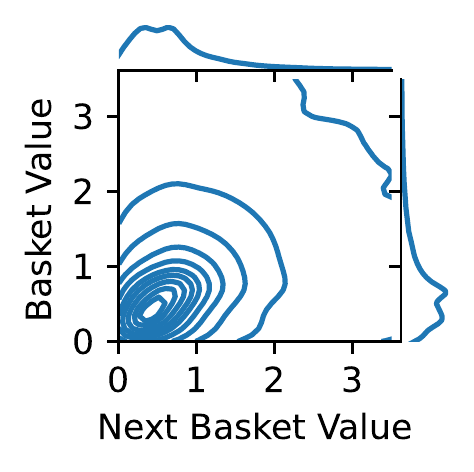}
    \includegraphics[height=0.17\textwidth, trim = 10mm 2mm 2mm 2mm, clip]{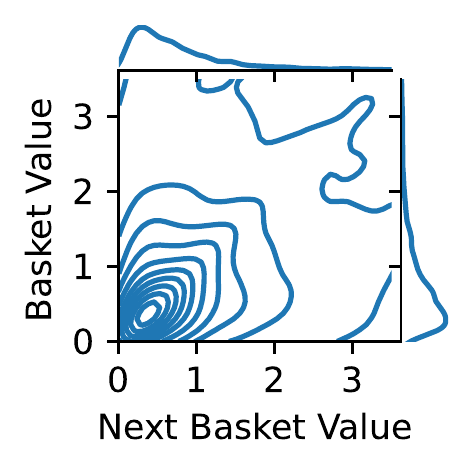}\\
    \includegraphics[height=0.17\textwidth, trim = 0mm 2mm 2mm 2mm, clip]{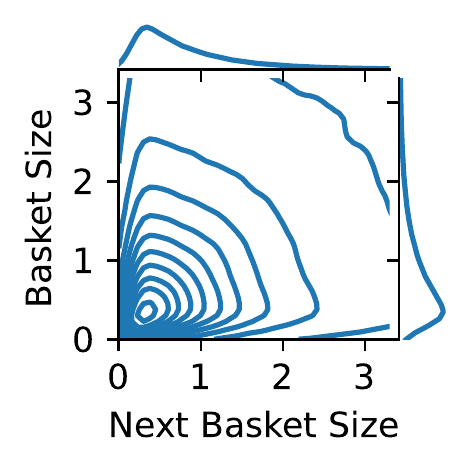}
    \includegraphics[height=0.17\textwidth, trim = 10mm 2mm 2mm 2mm, clip]{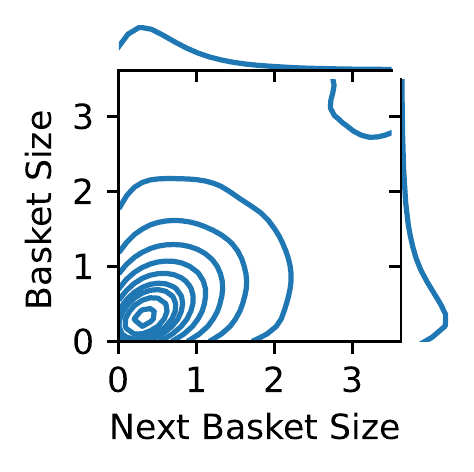}
    \includegraphics[height=0.17\textwidth, trim = 10mm 2mm 2mm 2mm, clip]{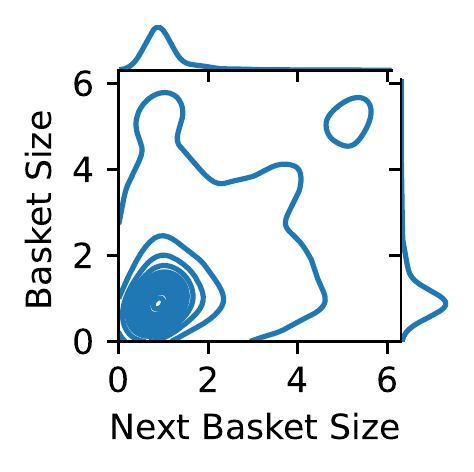}\\
  \end{center}
  \vspace*{-6pt}
  \caption{Kernel density estimation plots showing a user's transaction (Top row) value / (Bottom row) size against that of their next transaction---if it exists---in the (Left) ASOS, (Middle) UCI Online Retail II, and (Right) Olist Brazilian e-Commerce datasets. We normalize the transaction values/sizes by the ABV/ABS of respective datasets.}
  \Description[Six kernel density estimation plots (arranged in two by three) showing the size/value of an order is likely to be similar to that of the next order.]{Six kernel density estimation plots (arranged in two by three) showing the size/value of an order is likely to be similar to that of the next order. All plots have outlines resembling stingrays' undersides. The "head" is at the top-right corner, and the "wings" are on the top-left and bottom-right corners. There are also around ten contour lines near the bottom-left corner concentrating around the x equals to y diagonal. Both axes go from zero to just over three, except the bottom-right plot, which goes from zero to just over six.}
  \label{fig:abv_EDA_BasketValue_True}
\end{figure}

For each of the datasets, we plot the value and size of a user's transaction against that of their next transaction (if it exists) in Figure~\ref{fig:abv_EDA_BasketValue_True}. We observe from the figures a high empirical density along the $x=y$ diagonal and strong directionality in the kernel density estimates. This suggests a positive correlation between transaction values/sizes from the same user.
% Plot for value/selling price of items not shown but it is obvious there will be a correlation
For brevity, we do not show the plots for item values, yet contend a perfect correlation arises when a customer purchases multiple units of the same product.

As responses from the same user become correlated, we have $\textrm{Cov}(X_i, X_j) > 0$ for some~$i$ and $j$ and thus a higher SE than the vanilla sample SE ($\sqrt{\sigma^2/n}$ or its unbiased estimate $\sqrt{s^2/n}$).

% \footnote{It goes without saying that transactions (items) belonging to users who made only one transaction (bought only one item) during the experiment are considered independent to every other transaction (item) in practice, in the same spirit as how each user are considered independent to other users in A/B tests that split by user.}

\subsection{Re-estimating the SE using bootstrap}

\begin{figure}
  \begin{center}
    \includegraphics[width=0.475\textwidth, trim = 2mm 2mm 2mm 2mm, clip]{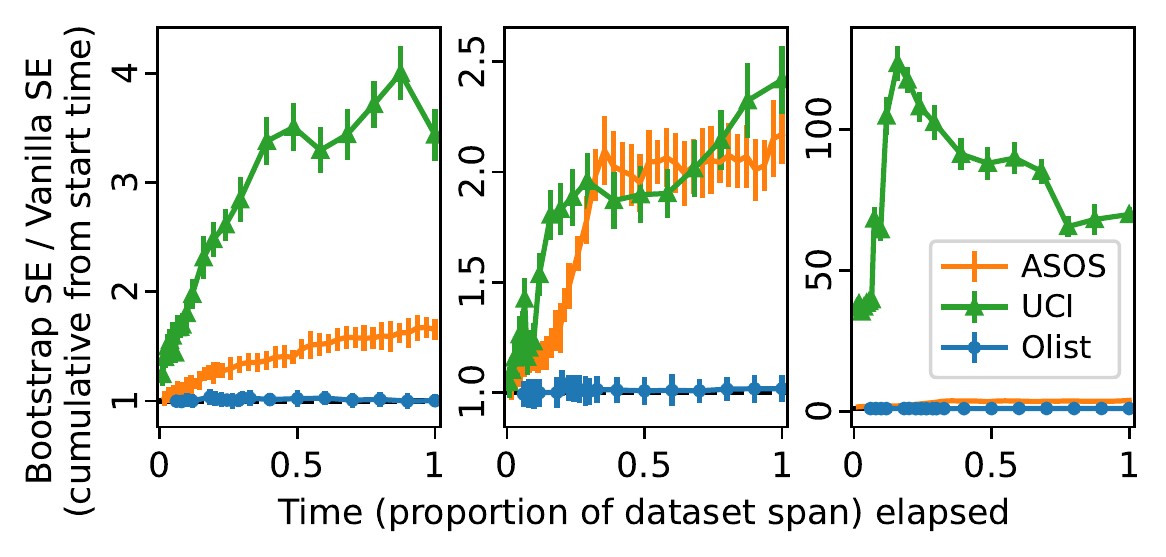}
  \end{center}
  \vspace*{-8pt}
  \caption{The ratio between the one-way bootstrap standard error (SE) estimates~\cite{bakshy13uncertainty} and the vanilla sample SE estimates for (Left) ABV, (Middle) ABS, and (Right) ASP. The error bars represent the 95\% confidence interval for the SE estimates.}
  \Description[Three line plots showing different ratios for different datasets and metrics at different times.]{Three line plots showing different ratios for different datasets and metrics at different times. Each plot contains an orange solid (marked "ASOS"), a green with triangles (marked "UCI"), a blue with dots (marked "Olist") line, and a black dashed line at Y equals one. The x-axis reads "Time (proportion of dataset span) elapsed" and goes from zero to one for all three plots. In the leftmost plot, the green line starts from just over one on the left, rises to three at one-third to the right, and fluctuates between three and four to the right; the orange line rises steadily from one to around 1.5 from left to right; the blue line stays at one from left to right, covering the black-dashed line. In the middle plot, the green line starts from just over one on the left, rises and stays at just below two in the middle half of the plot, and rises to 2.5 on the right; the orange line starts from one on the left, rises to two one-third to the right and stays there; the blue line stays at one from left to right, covering the black-dashed line. Finally, in the rightmost plot, the green line starts at around 35 on the left, shoots up to around 125 one-fifth to the right, decreases with minor fluctuation, and ends at around 65 on the right; the orange and blue lines are practically flat throughout near zero.}
  \label{fig:abv_SE_ABV}
\end{figure}

We then explore how much the SE can inflate. We apply the bootstrap procedure described in Section~2.2 of~\cite{bakshy13uncertainty} to the datasets described above to re-estimate the SE. 
Bootstrapping generates a resample by sampling the original set of responses with replacement (or applying a random weight in this case). This yields a different sample mean. By repeating the process many times and taking many bootstrap means, we can estimate the SE by calculating the standard deviation of the bootstrap.

We use a one-way bootstrap
% (a.k.a. block bootstrap)
to account for the dependency between transactions/items and users. Instead of generating the resample by sampling each transaction/item individually, we sample clusters of transactions/items belonging to the same user, mirroring our randomization process.\footnote{For ASP, in addition to the dependency between items and users, there may be additional dependence between items and products/stock-keeping units (SKUs). One may account for both using a two-way bootstrap described in Section 2.3 of~\cite{bakshy13uncertainty}.}
We use random weights generated from Poisson(1) to reweight our samples before calculating a bootstrap mean and estimating the SE from 500 bootstrap means.

% Figure~\ref{fig:abv_SE_ABV} shows how the one-way bootstrap SE differs from the vanilla sample SE in successive expanding windows, corresponding to different potential A/B test duration. We observe the bootstrap SE is significantly higher than the vanilla sample SE, with the difference being roughly 1.40x for ABV, 1.95x for ABS, and 3.70x for ASP during the first 30 days within the ASOS dataset.

Figure~\ref{fig:abv_SE_ABV} shows how the one-way bootstrap SE differs from the vanilla sample SE in successive expanding windows, corresponding to different potential A/B test duration. We observe the bootstrap SE is significantly higher than the vanilla sample SE, with the exact ratio between the one-way bootstrap SE and vanilla sample SE heavily dependent on the dataset/audience, business metric, and duration. For example, the bootstrap SE for ASP in the UCI Online Retail~II dataset is >100x the vanilla SE as many of their users buy tens or hundreds of the same product. Meanwhile, the ratios for ABV and ABS in the Olist Brazilian e-Commerce dataset are practically one. This is due to only 3.2\% of transactions coming from returning users. Given such differences, experimenters should strive to obtain a reasonably accurate estimate.

\begin{figure}
    \begin{center}
    \includegraphics[width=0.48\textwidth, trim = 2mm 2mm 2mm 2mm, clip]{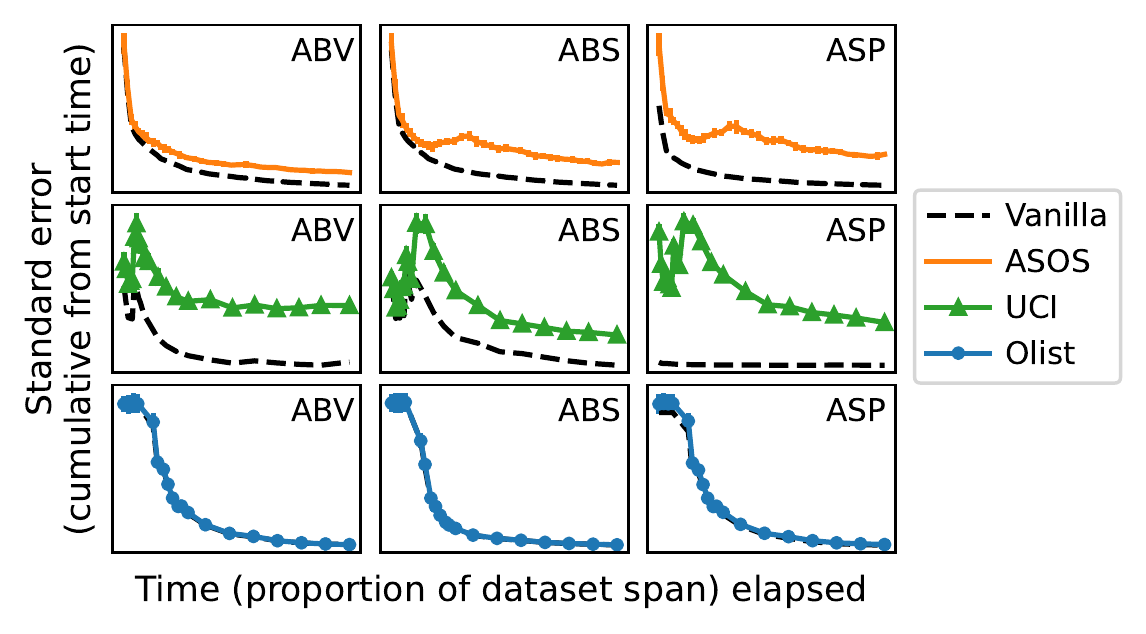}
    \end{center}
    \vspace*{-8pt}
    \caption{Trajectories of the one-way bootstrap standard error (SE) estimates  (solid colored lines) and the vanilla SE estimates (dashed black lines) under different dataset-metric combinations. Estimates are cumulative from day one of the datasets. All x- and y-axes are on different scales.}
    \Description[Nine line plots (arranged in three by three), each with two lines showing different trajectories in standard error estimates against time elapsed for different dataset-metric combinations.]{Nine line plots (arranged in three by three), each with two lines showing different trajectories in standard error estimates against time elapsed for different dataset-metric combinations. All plots contain a black dashed line marked "Vanilla". The plots at the top, middle, and bottom rows each contain an orange solid (marked "ASOS"), a green with triangles (marked "UCI"), and a blue with dots (marked "Olist") line, respectively. In addition, the plots in the left, middle, and right columns are marked "ABV", "ABS", and "ASP", respectively. The top-left plot shows the two lines in an L-shaped trajectory, with the orange line above the black line. The top-middle and top-right plots show similar L-shaped trajectories, with a bump in the orange line one-third to the right. The middle-left plot shows the green and black lines starting from the top-left, dropping, surging, and dropping again within the left one-fifth space, then holding flat to the right; the green line holds flat at half height while the black line holds flat at the bottom. The middle-middle plot shows lines with similar initial trajectories as the middle-left plot, though both lines decrease parallelly to the bottom-right corner. The middle-right plot shows a green line with a trajectory similar to the middle-middle plot and a flat black line at the bottom. Each of the bottom row plots shows the blue line overlapping the black line, which starts at the top-left corner and stays flat, drops quickly at one-fifth to the right, and decreases steadily from one-quarter from the bottom to the bottom-right corner.}
    \label{fig:abv_svbse_trajectory}
\end{figure}

We also observe from Figure~\ref{fig:abv_svbse_trajectory} that the bootstrap SE may no longer drop as more transactions are involved over time and may go up again in some cases. This is likely due to returning users making further transactions and thus creating more and larger clusters of transaction/item values and sizes. It leads to the response variance increasing, sometimes more quickly than the increase in sample size. The observation suggests lengthening an experiment solely to collect more responses (hence lowering the~SE) may backfire. In addition to established practices on sample size estimation~\cite{richardson22bayesian}, experimenters should also consider how the variance of their metrics evolves when designing A/B tests with dependent data.

\subsection{Impact on A/B test decisions}
\label{sec:abv_power_CI_impact}

We finally discuss how an inflated SE due to dependent data can affect decisions made from a null hypothesis significance test. Firstly, it reduces the power of the test, making any potential treatment effect harder to detect. Consider a two-tailed Student's $t$-test with significance level~$\alpha$ and $\nu$ d.f. Its power is
\begin{align}
     1 - T_{\nu}\big(t_{\nu, 1 - \alpha/2} - \theta/ \textrm{SE}\big) + T_{\nu}\big(-t_{\nu, 1 - \alpha/2} - \theta / \textrm{SE}\big),
     \label{eq:abv_test_power}
\end{align}
where $\theta$ is the effect size, $T_{\nu}(\cdot)$ is the CDF, and $t_{\nu, q}$ is the $q^{\textrm{th}}$ quantile of a $t$-distributed r.v. If the SE increases, then both the standardized effect size $\theta / \textrm{SE}$ and the test power in Expression~\eqref{eq:abv_test_power} decrease.

Secondly, having an inflated SE without knowing so will lead to tests producing confidence intervals that are too narrow, risking more false positives. In the test above, the $(1-\alpha)$ confidence interval~(CI) is 
$\big[\bar{x} \,\pm\, t_{\nu, 1 - \alpha/2} \cdot \textrm{SE}\big]$.
Fixing the CI bounds, we observe that $t_{\nu, 1 - \alpha/2}$ must decrease (and cease to be a $(1 - \alpha/2)$~quantile) when the SE increases. This means the said CI will no longer have a $(1-\alpha)$ but a lower coverage, i.e., there is now a lower chance the interval will contain the actual effect size.

\begin{figure}
  \begin{center}
    \includegraphics[width=0.23\textwidth, trim = 1mm 2mm 2mm 2mm, clip]{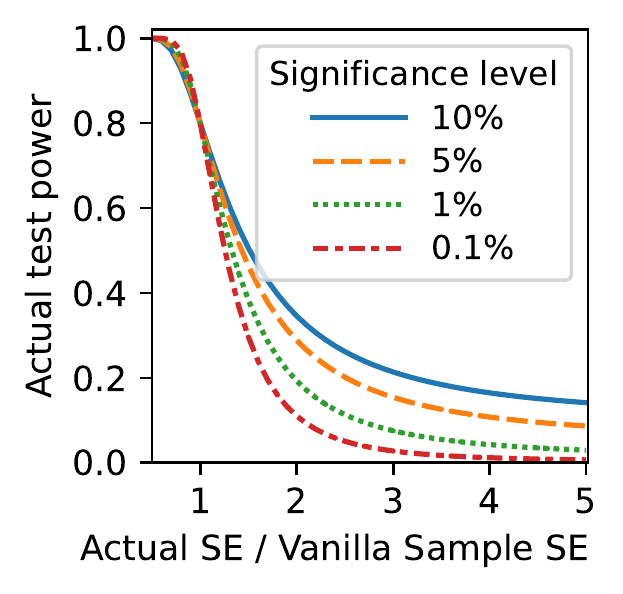}
    \includegraphics[width=0.23\textwidth, trim = 1mm 2mm 2mm 2mm, clip]{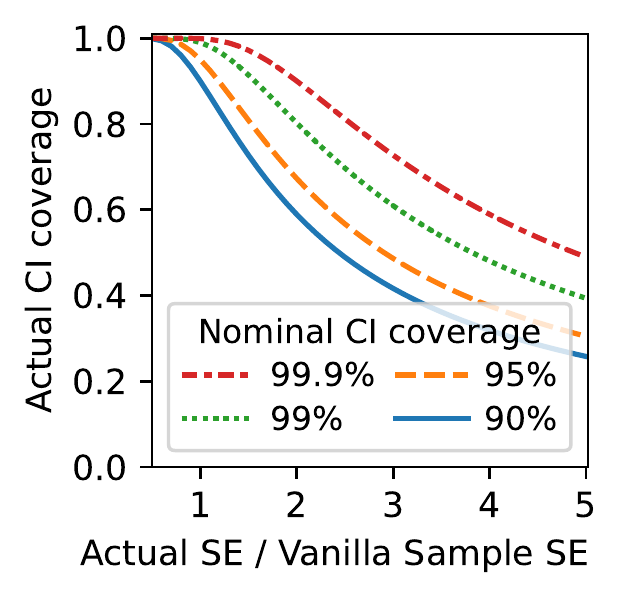} 
  \end{center}
  \vspace*{-8pt}
  \caption{(Left) The actual test power of and (Right) the coverage of the centered confidence interval (CI) from a two-tailed $z$-test against different standard errors (SE), expressed as a multiple of the vanilla sample SE $\sqrt{s^2/n}$. We choose the test parameters such that the test power $=80\%$ and the CI coverage $=$ the nominal values when the SE multiples are one.}
  \Description[Two line plots, side-by-side, each with four coloured lines showing the actual test power and confidence interval coverage decreases with increasing standard error.]{Two line plots, side-by-side, each with four coloured lines showing the actual test power and confidence interval coverage decreases with increasing standard error. The four lines in the left plot are, from top to bottom, blue solid (marked "Significance level 10\%"), orange dashed (marked "5\%"), green dotted (marked "1\%"), and red dot-dashed (marked "0.1\%"). The four lines in the right plot are, from top to bottom, red dot-dashed (marked "Nominal CI coverage 99.9\%"), green dotted (marked "99\%"), orange dashed (marked "95\%"), and blue solid (marked "90\%"). All lines in the left plot resemble a vertically-flipped sigmoid curve, starting near y equals one when x is less than one, dropping when x is between one and two, and decreasing steadily afterwards to the right. The top blue line in the left plot ends near Y equals 0.2, and the red dot-dashed line ends near Y equals zero. All lines in the right plot also resemble a vertically-flipped sigmoid curve but decrease more slowly than the left plot, starting near y equals one when x is less than one. The top red dot-dashed line ends near Y equals 0.5 on the right, and the bottom blue solid line ends near Y equals 0.3 on the right.}
  \label{fig:abv_actual_test_power_CI_coverage}
\end{figure}

To show the full extent of the issues above, we plot the test power and CI coverage of a two-tailed $z$-test, essentially a Student's $t$-test with large degrees of freedom, against different SEs in Figure~\ref{fig:abv_actual_test_power_CI_coverage}. We observe that the power of a $z$-test with a 5\% significance level (dashed line) tumbles from 80\% to around~29\% when we merely double the vanilla sample SE. Moreover, the centered 95\% CI calculated using the vanilla sample SE would only have roughly 67\% coverage. These results reinforce the importance of having an accurate estimate of the SE.

% We provide the quantities under different commonly used significance levels and hence nominal CI coverage.

\section{Lessons Learned During Deployment}

We finally share some lessons learned when incorporating the calculations of e-commerce metrics into ASOS.com's experiment analysis platform, which has been in production since September 2019. The analysis platform extends the functionality of a third-party experimentation platform, namely splitting, logging, and alerting. It adds value to the business by computing and performing statistical tests on bespoke business metrics that use internal datasets.
% It feature data pipelines that (1) extract metadata, (2) preprocess logs from the experimentation platform and multiple internal data sources into common format tables, (3) compute metrics from the tables, and (4) perform statistical tests and calculated associated quantities.

As mentioned in Section~1, experimenters in digital technology prefer the use of bootstrap (see Section~2.2 of~\cite{bakshy13uncertainty}) and the delta method (see Section~3 of~\cite{deng18applying}) to re-estimate the SE, which acts as a plug-in estimate for subsequent statistical tests. We described and implemented the bootstrap approach in Section 2.1 when we quantify the extent of the SE inflation.
The delta method approach expresses a transaction-/item-based metric as the quotient of two user-based metrics. For example, we can express ABV as the quotient of the business metrics ``mean spend per user (across all baskets)'' and ``mean number of baskets per user.'' Given both metrics in the quotient are asymptotically normal and based on randomization unit-based responses, which are commonly assumed to be i.i.d.~\cite{deng17trustworthy}, we can use the delta method to estimate the variance of the quotient. The resultant formula has an explicit term that accounts for the dependency between users and transactions/items.

\paragraph{Delta method - a straightforward formula with hidden complexity and inflexibility}
The delta method approach is more data efficient than the bootstrap approach. It requires only group-level statistics instead of analysis/sub-randomization unit-level responses required by bootstrap methods~\cite{liu2021datasets}. However, it is unclear how the results in~\cite{deng18applying}, developed with a one-way dependency between randomization and analysis units in mind, would apply to business metrics that feature responses dependent on other types of units.\footnotemark[6] The approach also requires considerable statistical mastery to understand and implement, risking decreased engagement from engineering teams and non-technical stakeholders due to general apprehensiveness towards ``black box'' algorithms.

\paragraph{Multi-way bootstrap may be overkill by being overly conservative}
The multi-way bootstrap described in Section 2.3 of~\cite{bakshy13uncertainty} aims to address dependencies between different types of units.
% \footnote{For ASP, a two-way bootstrap assigns one random weight to each user and one random weight to each product, before multiplying the associated weights to determine the final weighting of an item in each bootstrap resample. This differs from a one-way bootstrap, which assigns the same random weight to each user and all their items.}
However, we find the SE estimates for item-based metrics (e.g., ASP) from a multi-way bootstrap impractically large (see Table~\ref{tab:ecomm_asp_oneway_twoway_comparison}). This appears to match the results reported in~\cite{bakshy13uncertainty}, where two-way bootstrap SE estimates generally yield CIs there are overly conservative in A/A tests. While the true SE will always be unknown, we believe overestimating it is just as bad as underestimating it, as experimenters will then struggle to design any experiments with sufficient power.

We currently use a one-way, user-based bootstrap to estimate the SE of e-commerce metrics on our analysis platform. The approach mimics the randomization process, and there is extensive evidence (both in~\cite{bakshy13uncertainty} and within the organization) that it produces CIs with the right coverage in A/A tests.
With that said, we are constantly evaluating whether a multi-way bootstrap will become necessary, especially for experiments with product-treatment interactions.\footnote{Section 4 of~\cite{bakshy13uncertainty} shows a one-way, user-based bootstrap may underestimate the SE of item-based metrics when there are product-treatment interactions. Note ``products'' in this paper are referred to as ``items'' in~\cite{bakshy13uncertainty} due to different business contexts.}

\begin{table}
  \caption{The ratio between the bootstrap standard error (SE) estimates and the vanilla sample SE estimates for ASP. Each entry represents the mean and standard deviation of the ratios across multiple experiment runs.}
  \label{tab:ecomm_asp_oneway_twoway_comparison}
  \vspace*{-6pt}
  \begin{tabular}{cccc}
    \toprule
    Dataset & Duration & One-way bootstrap & Two-way bootstrap\\
    \midrule
    ASOS &
    First 30d &
    3.51 $\pm$ 0.13 &
    17.77 $\pm$ 0.56 \\
    \midrule
    \multirow{3}{*}{UCI} &
    First 10\%&
    64.45 $\pm$ 1.94 &
    72.83 $\pm$ 2.20 \\
    &
    First 50\% &
    87.95 $\pm$ 2.82 &
    105.15 $\pm$ 3.33 \\
    &
    All &
    69.82 $\pm$ 1.37 &
    94.33 $\pm$ 2.28 \\
    \midrule
    \multirow{3}{*}{Olist} &
    First 20\% &
    1.104 $\pm$ 0.035 &
    2.083 $\pm$ 0.075 \\
    &
    First 50\% &
    1.058 $\pm$ 0.036 &
    2.531 $\pm$ 0.066 \\
    &
    All &
    1.076 $\pm$ 0.037 &
    3.364 $\pm$ 0.109 \\
    \bottomrule
  \end{tabular}
\end{table}

\paragraph{Further consideration - error v.s. runtime/cost tradeoff}
Bootstrap methods introduce their error (more precisely, \emph{variance component}) to SE estimates during the resampling process. This is in addition to the (inflated) sampling uncertainty and generally scales along $O\big(1/\sqrt{B}\big)$, where $B$ is the number of bootstrap resamples~\cite{efron93introduction}. Given the runtime of bootstrap methods scales along~$O(B)$, at some point, it is no longer economically practical for a production system to draw further resamples to reduce the error to SE estimates. 

Deciding where the tipping point is more an art than science and depends more on individual needs. Unlike many technology giants which perform analysis in a streaming fashion, our analysis platform features batch-processing pipelines due to temporal constraints from upstream data sources. As a result, our priority is to keep the processing jobs' runtime (and hence compute cost) low. We find $B \in [500, 1000]$ works well for us as it leads to a coefficient of variation (error of SE estimate divided by mean SE estimate) of~<5\%. Having that said, we believe the tradeoff also applies to streaming systems as the compute latency increases with a larger~$B$, which impacts system availability.

\section{Conclusion}

A/B tests with dependent data are common in e-commerce, especially when experimenters are randomizing by users and analyzing transactions/items as they measure changes to business metrics such as ABV, ABS, and ASP. However, the risk they pose, namely experimenters obtaining an incorrect (usually a lower) standard error (SE) estimate using the vanilla sample SE formula, is often not adequately mitigated. We attribute the phenomenon to the lack of related evidence, which leads to insufficient awareness from practitioners in e-commerce.

In this paper, we discussed why the vanilla SE estimate could be incorrect, demonstrated the magnitude of the problem using three e-commerce transactions datasets, and outlined its impact on A/B test outcomes. We did not offer any new solution to the problem as many viable mitigating approaches already exist. Instead, we critiqued popular approaches in digital technology and discussed practical considerations when implementing these approaches in an experiment analysis platform. We intend for this paper to be a short but crucial piece of evidence that practitioners can utilize to justify introducing said mitigating approaches into online experiment design and analysis processes within their organization.

% \begin{acks}
% The research is part-funded by the EPSRC CDT in Modern Statistics and Statistical Machine Learning at Imperial College London and University of Oxford (StatML.IO) and ASOS.com.
% \end{acks}

\bibliographystyle{ACM-Reference-Format}
\bibliography{abv_references}


\begin{thebibliography}{15}

%%% the Web address.

\ifx \showCODEN    \undefined \def \showCODEN     #1{\unskip}     \fi
\ifx \showDOI      \undefined \def \showDOI       #1{#1}\fi
\ifx \showISBNx    \undefined \def \showISBNx     #1{\unskip}     \fi
\ifx \showISBNxiii \undefined \def \showISBNxiii  #1{\unskip}     \fi
\ifx \showISSN     \undefined \def \showISSN      #1{\unskip}     \fi
\ifx \showLCCN     \undefined \def \showLCCN      #1{\unskip}     \fi
\ifx \shownote     \undefined \def \shownote      #1{#1}          \fi
\ifx \showarticletitle \undefined \def \showarticletitle #1{#1}   \fi
\ifx \showURL      \undefined \def \showURL       {\relax}        \fi
% invisible to TeX
\providecommand\bibfield[2]{#2}
\providecommand\bibinfo[2]{#2}
\providecommand\natexlab[1]{#1}
\providecommand\showeprint[2][]{arXiv:#2}

\bibitem[Baayen et~al\mbox{.}(2008)]%
        {baayen08mixedeffects}
\bibfield{author}{\bibinfo{person}{R.~H. Baayen}, \bibinfo{person}{D.~J.
  Davidson}, {and} \bibinfo{person}{D.~M. Bates}.}
  \bibinfo{year}{2008}\natexlab{}.
\newblock \showarticletitle{Mixed-effects modeling with crossed random effects
  for subjects and items}.
\newblock \bibinfo{journal}{\emph{Journal of Memory and Language}}
  \bibinfo{volume}{59}, \bibinfo{number}{4} (\bibinfo{year}{2008}),
  \bibinfo{pages}{390--412}.
\newblock
\showISSN{0749-596X}
\urldef\tempurl%
\url{https://doi.org/10.1016/j.jml.2007.12.005}
\showDOI{\tempurl}
\newblock
\shownote{Special Issue: Emerging Data Analysis}.


\bibitem[Bakshy and Eckles(2013)]%
        {bakshy13uncertainty}
\bibfield{author}{\bibinfo{person}{Eytan Bakshy} {and} \bibinfo{person}{Dean
  Eckles}.} \bibinfo{year}{2013}\natexlab{}.
\newblock \showarticletitle{Uncertainty in Online Experiments with Dependent
  Data: An Evaluation of Bootstrap Methods}. In
  \bibinfo{booktitle}{\emph{Proceedings of the 19th ACM SIGKDD International
  Conference on Knowledge Discovery and Data Mining}} (Chicago, Illinois, USA)
  \emph{(\bibinfo{series}{KDD '13})}. \bibinfo{publisher}{Association for
  Computing Machinery}, \bibinfo{address}{New York, NY, USA},
  \bibinfo{pages}{1303–1311}.
\newblock
\showISBNx{9781450321747}
\urldef\tempurl%
\url{https://doi.org/10.1145/2487575.2488218}
\showDOI{\tempurl}


\bibitem[Cameron et~al\mbox{.}(2011)]%
        {cameron11robust}
\bibfield{author}{\bibinfo{person}{A.~Colin Cameron}, \bibinfo{person}{Jonah~B.
  Gelbach}, {and} \bibinfo{person}{Douglas~L. Miller}.}
  \bibinfo{year}{2011}\natexlab{}.
\newblock \showarticletitle{Robust Inference With Multiway Clustering}.
\newblock \bibinfo{journal}{\emph{Journal of Business \& Economic Statistics}}
  \bibinfo{volume}{29}, \bibinfo{number}{2} (\bibinfo{year}{2011}),
  \bibinfo{pages}{238--249}.
\newblock
\urldef\tempurl%
\url{https://doi.org/10.1198/jbes.2010.07136}
\showDOI{\tempurl}
\showeprint{https://doi.org/10.1198/jbes.2010.07136}


\bibitem[Chen et~al\mbox{.}(2012)]%
        {chen12datamining}
\bibfield{author}{\bibinfo{person}{Daqing Chen}, \bibinfo{person}{Sai~Laing
  Sain}, {and} \bibinfo{person}{Kun Guo}.} \bibinfo{year}{2012}\natexlab{}.
\newblock \showarticletitle{Data mining for the online retail industry: A case
  study of RFM model-based customer segmentation using data mining}.
\newblock \bibinfo{journal}{\emph{Journal of Database Marketing {\&} Customer
  Strategy Management}} \bibinfo{volume}{19}, \bibinfo{number}{3}
  (\bibinfo{date}{01 Sep} \bibinfo{year}{2012}), \bibinfo{pages}{197--208}.
\newblock
\showISSN{1741-2447}
\urldef\tempurl%
\url{https://doi.org/10.1057/dbm.2012.17}
\showDOI{\tempurl}


\bibitem[Deng et~al\mbox{.}(2018)]%
        {deng18applying}
\bibfield{author}{\bibinfo{person}{Alex Deng}, \bibinfo{person}{Ulf Knoblich},
  {and} \bibinfo{person}{Jiannan Lu}.} \bibinfo{year}{2018}\natexlab{}.
\newblock \showarticletitle{Applying the Delta Method in Metric Analytics: A
  Practical Guide with Novel Ideas}. In \bibinfo{booktitle}{\emph{Proceedings
  of the 24th ACM SIGKDD International Conference on Knowledge Discovery \&
  Data Mining}} (London, United Kingdom) \emph{(\bibinfo{series}{KDD '18})}.
  \bibinfo{publisher}{Association for Computing Machinery},
  \bibinfo{address}{New York, NY, USA}, \bibinfo{pages}{233–242}.
\newblock
\showISBNx{9781450355520}
\urldef\tempurl%
\url{https://doi.org/10.1145/3219819.3219919}
\showDOI{\tempurl}


\bibitem[Deng et~al\mbox{.}(2017)]%
        {deng17trustworthy}
\bibfield{author}{\bibinfo{person}{Alex Deng}, \bibinfo{person}{Jiannan Lu},
  {and} \bibinfo{person}{Jonthan Litz}.} \bibinfo{year}{2017}\natexlab{}.
\newblock \showarticletitle{Trustworthy Analysis of Online {A/B} Tests:
  Pitfalls, Challenges and Solutions}. In \bibinfo{booktitle}{\emph{Proceedings
  of the Tenth ACM International Conference on Web Search and Data Mining}}
  (Cambridge, United Kingdom) \emph{(\bibinfo{series}{WSDM '17})}.
  \bibinfo{publisher}{Association for Computing Machinery},
  \bibinfo{address}{New York, NY, USA}, \bibinfo{pages}{641–649}.
\newblock
\showISBNx{9781450346757}
\urldef\tempurl%
\url{https://doi.org/10.1145/3018661.3018677}
\showDOI{\tempurl}


\bibitem[Dua and Graff(2019)]%
        {dua2019UCI}
\bibfield{author}{\bibinfo{person}{Dheeru Dua} {and} \bibinfo{person}{Casey
  Graff}.} \bibinfo{year}{2019}\natexlab{}.
\newblock \bibinfo{title}{{UCI} Machine Learning Repository}.
\newblock
\newblock
\urldef\tempurl%
\url{http://archive.ics.uci.edu/ml}
\showURL{%
\tempurl}


\bibitem[Eckles et~al\mbox{.}(2017)]%
        {eckles17design}
\bibfield{author}{\bibinfo{person}{Dean Eckles}, \bibinfo{person}{Brian
  Karrer}, {and} \bibinfo{person}{Johan Ugander}.}
  \bibinfo{year}{2017}\natexlab{}.
\newblock \showarticletitle{Design and Analysis of Experiments in Networks:
  Reducing Bias from Interference}.
\newblock \bibinfo{journal}{\emph{Journal of Causal Inference}}
  \bibinfo{volume}{5}, \bibinfo{number}{1} (\bibinfo{year}{2017}),
  \bibinfo{pages}{20150021}.
\newblock
\urldef\tempurl%
\url{https://doi.org/10.1515/jci-2015-0021}
\showDOI{\tempurl}


\bibitem[Efron and Tibshirani(1994)]%
        {efron93introduction}
\bibfield{author}{\bibinfo{person}{Bradley Efron} {and}
  \bibinfo{person}{Robert~J. Tibshirani}.} \bibinfo{year}{1994}\natexlab{}.
\newblock \bibinfo{booktitle}{\emph{An Introduction to the Bootstrap}}.
\newblock Number~57 in \bibinfo{series}{Monographs on Statistics and Applied
  Probability}. \bibinfo{publisher}{Chapman \& Hall/CRC},
  \bibinfo{address}{Boca Raton, Florida, USA}.
\newblock
\urldef\tempurl%
\url{https://doi.org/10.1201/9780429246593}
\showURL{%
\tempurl}


\bibitem[Gui et~al\mbox{.}(2015)]%
        {gui15network}
\bibfield{author}{\bibinfo{person}{Huan Gui}, \bibinfo{person}{Ya Xu},
  \bibinfo{person}{Anmol Bhasin}, {and} \bibinfo{person}{Jiawei Han}.}
  \bibinfo{year}{2015}\natexlab{}.
\newblock \showarticletitle{Network A/B Testing: From Sampling to Estimation}.
  In \bibinfo{booktitle}{\emph{Proceedings of the 24th International Conference
  on World Wide Web}} (Florence, Italy) \emph{(\bibinfo{series}{WWW '15})}.
  \bibinfo{publisher}{International World Wide Web Conferences Steering
  Committee}, \bibinfo{address}{Republic and Canton of Geneva, CHE},
  \bibinfo{pages}{399–409}.
\newblock
\showISBNx{9781450334693}
\urldef\tempurl%
\url{https://doi.org/10.1145/2736277.2741081}
\showDOI{\tempurl}


\bibitem[Kohavi et~al\mbox{.}(2020)]%
        {kohavi20trustworthy}
\bibfield{author}{\bibinfo{person}{Ron Kohavi}, \bibinfo{person}{Diane Tang},
  {and} \bibinfo{person}{Ya Xu}.} \bibinfo{year}{2020}\natexlab{}.
\newblock \bibinfo{booktitle}{\emph{Trustworthy Online Controlled Experiments:
  A Practical Guide to {A/B} Testing} (\bibinfo{edition}{1} ed.)}.
\newblock \bibinfo{publisher}{Cambridge University Press}.
\newblock
\showISBNx{9781108724265}


\bibitem[Liu et~al\mbox{.}(2021)]%
        {liu2021datasets}
\bibfield{author}{\bibinfo{person}{C.~H.~Bryan Liu},
  \bibinfo{person}{{\^A}ngelo Cardoso}, \bibinfo{person}{Paul Couturier}, {and}
  \bibinfo{person}{Emma~J. McCoy}.} \bibinfo{year}{2021}\natexlab{}.
\newblock \showarticletitle{Datasets for Online Controlled Experiments}. In
  \bibinfo{booktitle}{\emph{Proceedings of the Neural Information Processing
  Systems Track on Datasets and Benchmarks 1 (NeurIPS Datasets and Benchmarks
  2021)}}.
\newblock
\urldef\tempurl%
\url{https://datasets-benchmarks-proceedings.neurips.cc/paper/2021/file/274ad4786c3abca69fa097b85867d9a4-Paper-round2.pdf}
\showURL{%
\tempurl}


\bibitem[Liu et~al\mbox{.}(2020)]%
        {liu20whatisthevalue}
\bibfield{author}{\bibinfo{person}{C.~H.~Bryan Liu},
  \bibinfo{person}{Benjamin~Paul Chamberlain}, {and} \bibinfo{person}{Emma~J.
  McCoy}.} \bibinfo{year}{2020}\natexlab{}.
\newblock \showarticletitle{What is the Value of Experimentation and
  Measurement?: Quantifying the Value and Risk of Reducing Uncertainty to Make
  Better Decisions}.
\newblock \bibinfo{journal}{\emph{Data Science and Engineering}}
  \bibinfo{volume}{5} (\bibinfo{year}{2020}), \bibinfo{pages}{152--167}.
\newblock
\showISSN{23641541}
\urldef\tempurl%
\url{https://doi.org/10.1007/s41019-020-00121-5}
\showDOI{\tempurl}


\bibitem[Olist and Sionek(2018)]%
        {olist18braziliandataset}
\bibfield{author}{\bibinfo{person}{Olist} {and} \bibinfo{person}{André
  Sionek}.} \bibinfo{year}{2018}\natexlab{}.
\newblock \bibinfo{title}{Brazilian E-Commerce Public Dataset by {Olist}
  {[Dataset]}}.
\newblock
\newblock
\urldef\tempurl%
\url{https://doi.org/10.34740/KAGGLE/DSV/195341}
\showDOI{\tempurl}


\bibitem[Richardson et~al\mbox{.}(2022)]%
        {richardson22bayesian}
\bibfield{author}{\bibinfo{person}{Thomas~S. Richardson}, \bibinfo{person}{Yu
  Liu}, \bibinfo{person}{James Mcqueen}, {and} \bibinfo{person}{Doug Hains}.}
  \bibinfo{year}{2022}\natexlab{}.
\newblock \showarticletitle{A {Bayesian} Model for Online Activity Sample
  Sizes}. In \bibinfo{booktitle}{\emph{Proceedings of The 25th International
  Conference on Artificial Intelligence and Statistics}}
  \emph{(\bibinfo{series}{Proceedings of Machine Learning Research},
  Vol.~\bibinfo{volume}{151})}, \bibfield{editor}{\bibinfo{person}{Gustau
  Camps-Valls}, \bibinfo{person}{Francisco J.~R. Ruiz}, {and}
  \bibinfo{person}{Isabel Valera}} (Eds.). \bibinfo{publisher}{PMLR},
  \bibinfo{pages}{1775--1785}.
\newblock
\urldef\tempurl%
\url{https://proceedings.mlr.press/v151/richardson22a.html}
\showURL{%
\tempurl}


\end{thebibliography}

% \newpage

\appendix

\end{document}